\documentclass[twocolumn,runningheads]{svjour2}
\smartqed  
\usepackage{graphicx}
 \usepackage{mathptmx}      
\usepackage{amsmath} 
\journalname{Astrophysics and Space Science}
\begin{document}

\title{Internal absorption of gamma-rays in relativistic blobs of active galactic nuclei
}


\author{Julian Sitarek         \and
        Wlodek Bednarek  
}


\institute{J. Sitarek \at
	      Department of Experimental Physics, Univeristy of Lodz \\
              Tel.: +48-42-355655 \\
              \email{jusi@kfd2.fic.uni.lodz.pl }           
           \and
           W. Bednarek \at
	      Department of Experimental Physics, Univeristy of Lodz \\
              \email{bednar@fizwe4.fic.uni.lodz.pl}           
}

\date{Received: date / Accepted: date}

\maketitle

\begin{abstract}
We investigate the production of gamma-rays in the inverse Compton (IC) 
scattering process by leptons accelerated inside relativistic blobs 
in jets of active galactic nuclei. Leptons are injected homogeneously 
inside the spherical blob and initiate IC $e^\pm$ pair cascade 
in the synchrotron radiation (produced by the same population of 
leptons, SSC model), provided that the optical depth for gamma-rays 
is larger than unity. It is shown that for likely parameters 
internal absorption of gamma-rays has to be important. 
We suggest that new type of blazars might be discovered by the 
future simultaneous X-ray and $\gamma$-ray observations, showing peak emissions
in the hard X-rays, and in the GeV $\gamma$-rays. 
Moreover, the considered scenario might be also responsible for the orphan X-ray flares recently reported from BL Lac type active galaxies. 
\end{abstract}

\section{Introduction}
\label{intro}

The multiwavelength spectra of BL Lac type active galactic nuclei (AGNs) show usually 
two prominent broad bumps which peak in the hard X-ray and TeV $\gamma$-ray energy ranges. Most of the simultaneous observations argued for strong  correlation of emission in these two energy ranges on short time scales.
However, more recently the so called orphan flares, e.g. X-ray orphan flares (from Mrk 421, Rebillot et al. 2006) and also $\gamma$-ray orphan flares (from 1ES1959+650, Daniel et al. 2005), have been observed in which no correlation between hard X-ray and TeV $\gamma$-ray emission seems to be observed. We wonder if the appearance of the X-ray type orphan flares can be related to the internal absorption of TeV $\gamma$-rays in the low energy synchrotron radiation produced by the same population of relativistic electrons. 

\section{Optically thick homogeneous SSC model}
\label{model}

\begin{figure*}[t]
\centering
  \includegraphics[scale=0.8]{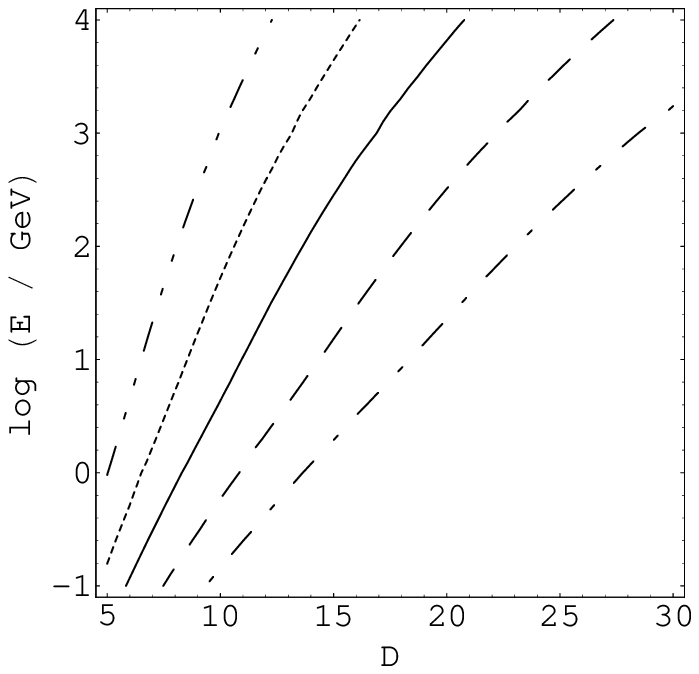}
  \includegraphics[scale=0.8]{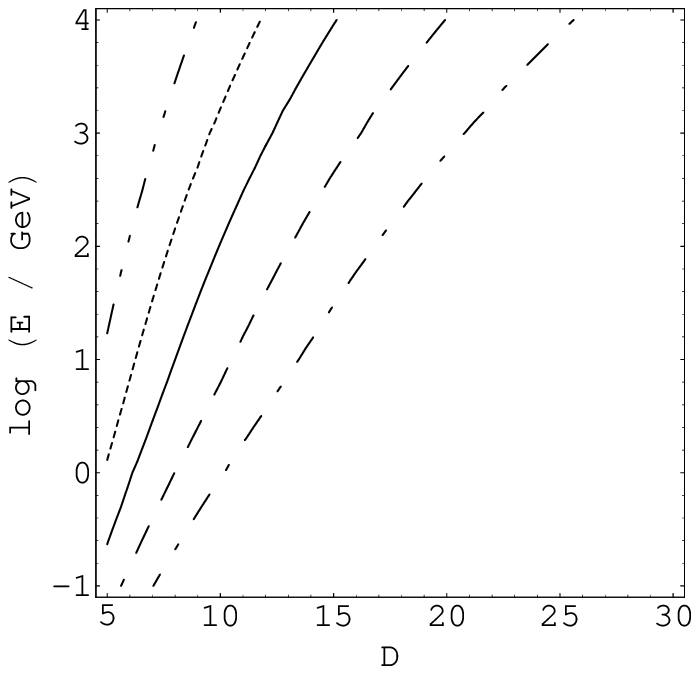}
\caption{The energies of $\gamma$-ray photons injected in the centre of the blob for which the optical depths in 
$\gamma-\gamma\rightarrow e^\pm$ collisions with the synchrotron radiation of the blob are equal to: $\tau=0.1$ (dot-dashed
curves), $\tau=0.3$ (dashed), $\tau=1$ (solid), $\tau=3$ (dotted), $\tau=10$ (dot-dot-dashed), are shown as a function of the Doppler factor of the blob and for
a few selected radii of the blob defined by the variability time scale $t_{\rm var} = 15$ min (left figure), and  1hr (right). These example calculations are performed for density and spectrum of synchrotron photons as derived for the flare on 16th April 1997 from Mrk 501.}
\label{fig1}       
\end{figure*}

Let us consider a simple scenario in which relativistic electrons fill uniformly
a spherical blob containing a random magnetic field. These electrons produce synchrotron
photons the spectra of which can extend up to X-ray energies.
Although the radiation field inside such an idealized blob is homogeneous, electrons at a specific location inside the blob (but outside the center of the blob) interact in fact with the anisotropic radiation. Electrons close to the border of the blob interact preferentially with the radiation coming from general direction of the central parts of the blob. 
Hence, also the IC $\gamma$-rays produced by these electrons are preferentially emitted  towards the central parts of the blob. For specific parameters of the 
blob, we calculate the optical depths, for electrons on the inverse Compton (IC) process, and for $\gamma$-rays, produced by these electrons, on $\gamma-\gamma\rightarrow e^\pm$ pair production process in collisions with synchrotron photons.  
As an example, we apply the X-ray observations of Mrk 501 observed during the 1997 April 16 flare (Catanese et al.~1997, Pian et al.~1997) and TeV
$\gamma$-ray observations between $\sim$200 GeV (Djannati-Atai et al. 1999) up to $\sim$20 TeV (Aharonian et al. 1999). It is assumed that the blob of relativistic electrons moves towards the observer with specific Doppler factor $D$. For the  parametrization of the X-ray spectrum and the derivation of the photon densities in the blob frame, see Bednarek \& Protheroe~(1999). The radius of the blob is estimated on the base of the reported variability time-scale $t_{\rm var}$ during the flare emission, $R=0.5c D t_{var}$.
The spectrum of the electrons in the blob can be derived from the observed synchrotron spectrum assumed in specific energy ranges to be simple power law. For the differential synchrotron photon spectrum, described in specific energy ranges by the general form $N_{\rm syn}\propto \epsilon^{-p}$ up to the maximum energies $\epsilon_{\max}$, we derive the spectrum of electrons in the blob, $N_{\rm e}\propto E^{-\alpha}$ up to energy $E_{\max}$.
A simple relation between  spectral indices of the synchrotron and electron spectra results from the basic features of the synchrotron radiation process, $\alpha = 2p-1$ and $E_{\rm max} =$ 
$(m c^2\epsilon_{\max} B_{\rm cr}/B)^{1/2}$, where $B$ is the magnetic field strength at the blob, and $B_{cr}=4.414\cdot10^{13}$ G (see e.g. Bednarek \& Protheroe~1999).

Relativistic electrons injected isotropically into the blob with such a spectrum can initiate IC $e^\pm$ pair cascade, provided that the emission region is compact enough (optical depths larger than unity). To find out for which blob parameters this is possible, we calculate the optical depths for $\gamma$-rays in the synchrotron radiation
of the blob for an arbitrary injection place of the $\gamma$-ray inside the blob.
Note that the electrons which gyrate in the magnetic field at the border of the blob will see the soft radiation anisotropic. They see the strongest radiation field from the direction of the centre of the blob.
Therefore, they produce TeV $\gamma$-ray photons in the IC process preferentially towards the centre of the blob since the probability of their interaction is the largest when they are instantaneously directed towards the centre. 
Produced TeV $\gamma$-rays move through the central regions of the blob, where the radiation field is the strongest. We take these anisotropic effects into account in our calculations. 

In Fig.~1, we show the optical depths for $\gamma-\gamma\rightarrow e^\pm$
pair production in collisions of IC $\gamma$-ray photons with synchrotron radiation as a function of the energies of $\gamma$-ray photons and the blob parameters (its dimension is defined by $t_{\rm var}$ and Doppler factor). It is clear that for blobs moving 
with Doppler factors estimated from the observations of the superluminal motions
in jets of AGNs, the  optical depths are above unity for $\gamma$-rays in the TeV energy range. For the shortest variability time scales reported in the case of Mrk 501 ($\sim$15 min), the optical depths for $\gamma$-rays with energies above $\sim$100 GeV in the blob rest frame are larger than unity already for Doppler factors lower than $\sim$15.
In such cases, absorption of TeV $\gamma$-rays in collisions with synchrotron photons have to be included in the calculations of the $\gamma$-ray spectrum emerging from the blobs.

\section{The cascade gamma-ray spectra}

\begin{figure*}[t]
\centering
  \includegraphics  [trim= 1 11 28 15, clip,scale=1.]{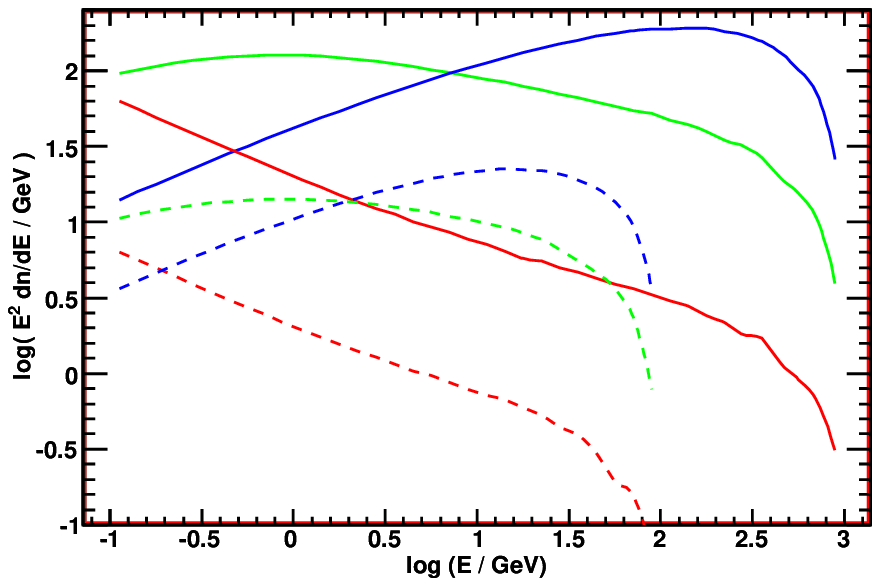}
    \includegraphics[trim=15 11 28 15, clip,scale=1.]{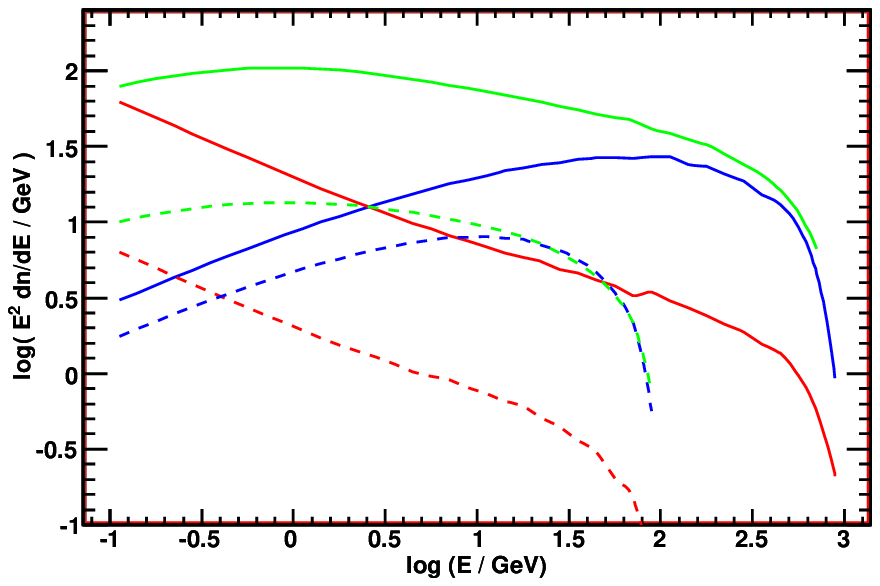} 
  \includegraphics  [trim= 1  2 28 15, clip,scale=1.]{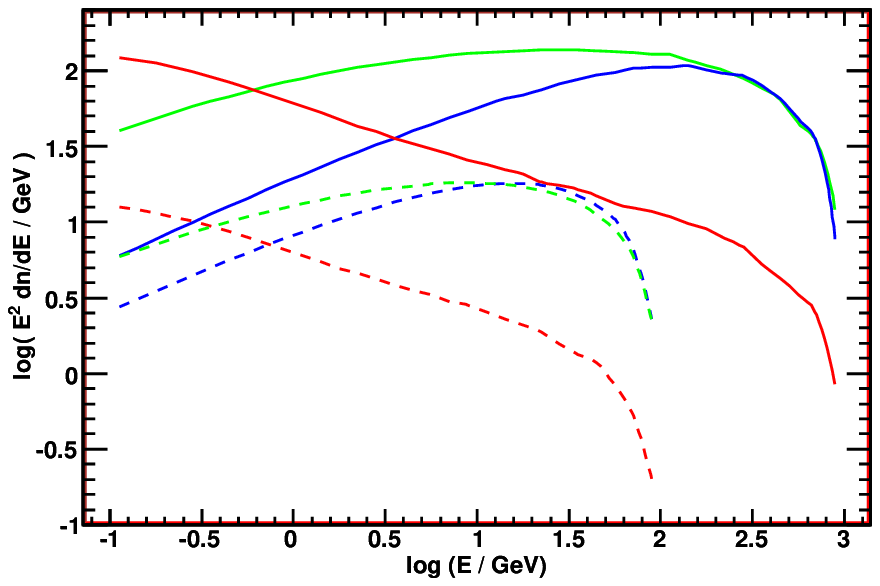}
    \includegraphics[trim=15  2 28 15, clip,scale=1.]{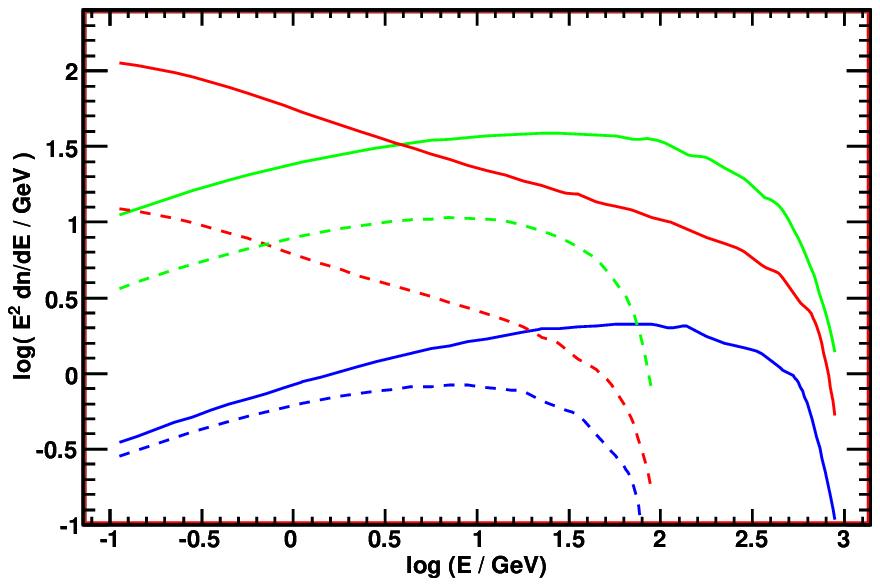}
\caption{Differential spectra of $\gamma$-ray photons multiplied by the energy squared (in the blob frame) which are produced in the 
optically thick SSC model by electrons with energies $E_{\rm e} = 0.1$TeV (dashed curves)
and 1 TeV (full) in the blob defined by following parameters:
blob radius determined by the variability time scales $t_{\rm var}=15$min (upper figures) and 1hr (lower), the magnetic field strength $B = 0.1$G (left figures) and 
1G (right), and different Doppler factors of the blob: $D=5$ (red curves),
10 (green), and 20 (blue).}
\label{fig:oneE}       
\end{figure*}

In order to obtain the $\gamma$-ray spectra escaping from the blob as a function of its
parameters, we assume that electrons, with the spectrum derived from the observed synchrotron spectrum during the flare from Mrk 501 observed during the 1997 April 16, are distributed homogeneously inside a blob. The blob is defined by its radius (determined by the variability time scale), the Doppler factor $D$, and the magnetic field strength $B$. These electrons cool on the synchrotron process and IC scattering of synchrotron photons. 
The IC $\gamma$-rays are produced by electrons anisotropically, preferentially toward
the direction of the strongest radiation field, i.e. toward the central parts of the blob.
These primary $\gamma$-rays can be absorbed in the blob synchrotron radiation, initiating the IC $e^\pm$ pair cascade. We follow the development of such anisotropic cascade (due to the
anisotropic radiation field), by applying the Monte Carlo method. The scenario discussed by us is in fact a modification of the synchrotron self-Compton model. However, the SSC model considered here is homogeneous with respect to injection of relativistic electrons, and the density of synchrotron photons, but inhomogeneous in respect to the synchrotron radiation field inside the blob, due to geometrical effects. We are interested  in 
situations in which absorption of $\gamma$-rays in collisions with synchrotron photons can  play an important role. Our IC $e^\pm$ pair cascade code also includes the synchrotron energy losses of secondary $e^\pm$ pairs.
The primary and secondary electrons in the blob are considered up to minimum energies
$E_{\rm min} = 100$ MeV. At first, we calculate the spectra of $\gamma$-ray escaping
from the blob in their own rest frame for monoenergetic electrons (normalized to one electron): 
$E_{\rm e} = 0.1$ TeV and $1$ TeV (see Fig.~2). As expected, the shape of these $\gamma$-ray spectra strongly depends on the Doppler factor of the blob, which is mainly responsible for the optical depths for $\gamma$-ray photons. For low Doppler factors, the spectra are
relatively steep peaking at GeV energies in the observer's frame (compare curves with different colours). In contrast, for large Doppler factors, the $\gamma$-ray spectra peak at the highest energy part, which is determined by the energies of injected electrons.
If the density of synchrotron photons in the blob is large (e.g. 
for small radius of the blob which occurs for relatively short variability time scale and small Doppler factors), then TeV $\gamma$-rays are absorbed and re-processed in the cascade into the GeV energy range (compare red, green and blue curves in the upper-left Fig.~2 or green curves in upper and bottom Fig.~2). 

\begin{figure*}[t]
\centering
  \includegraphics[trim= 1 11 28 15, clip,scale=1.]{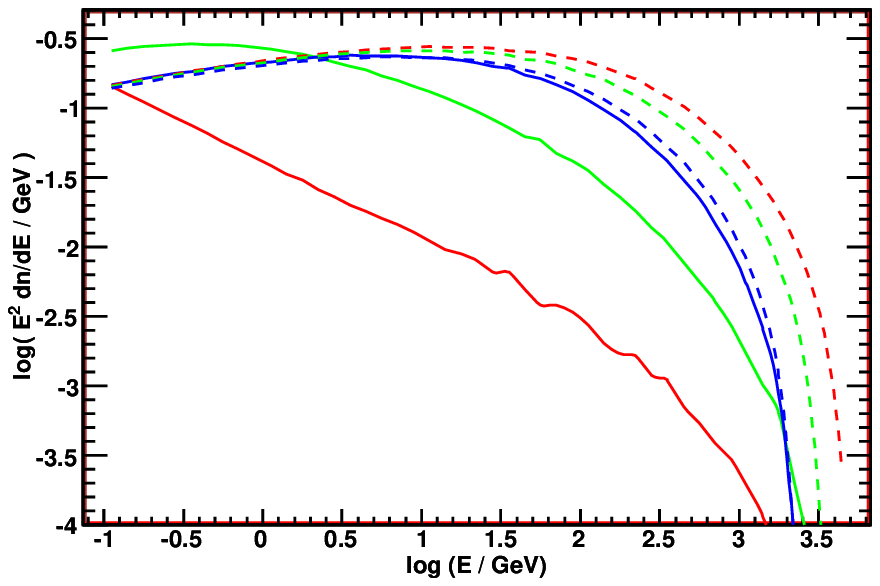}
  \includegraphics[trim=15 11 28 15, clip,scale=1.]{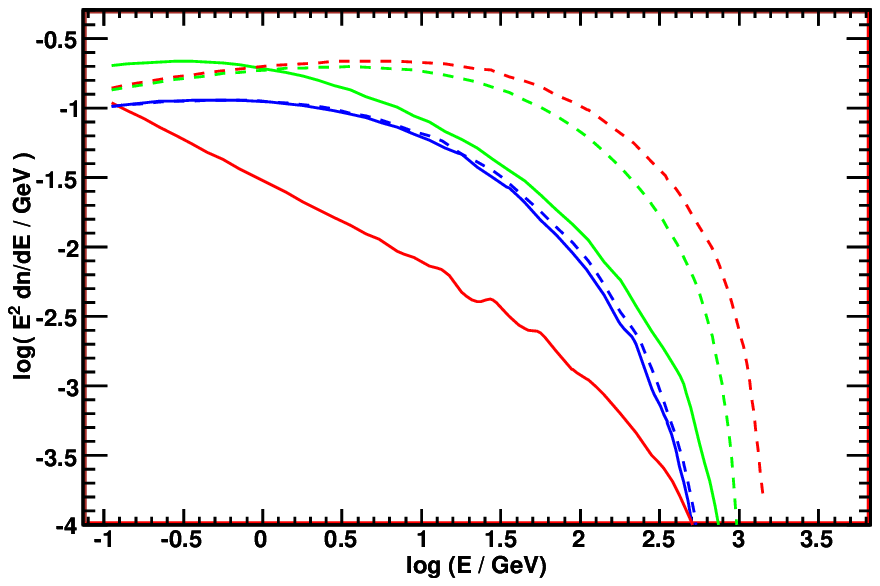} 
  \includegraphics[trim= 1  2 28 15, clip,scale=1.]{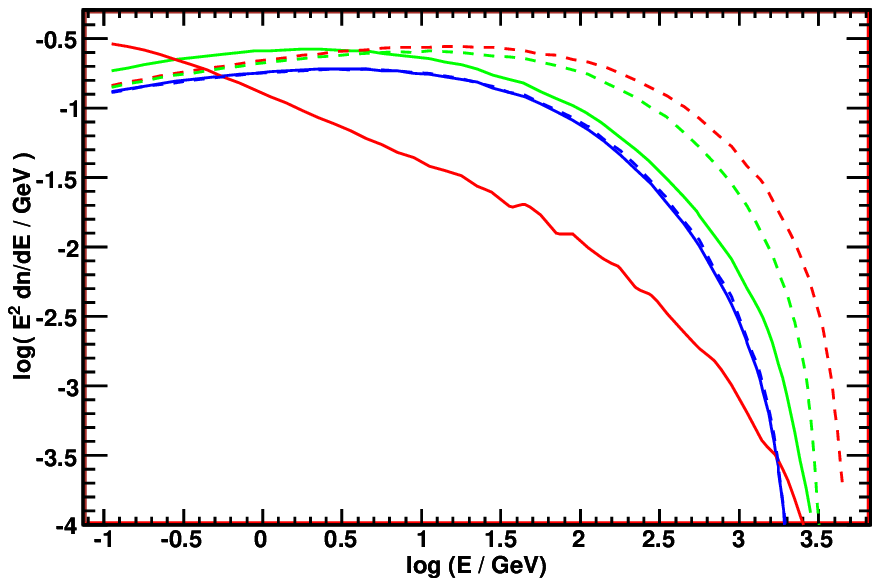}
  \includegraphics[trim=15  2 28 15, clip,scale=1.]{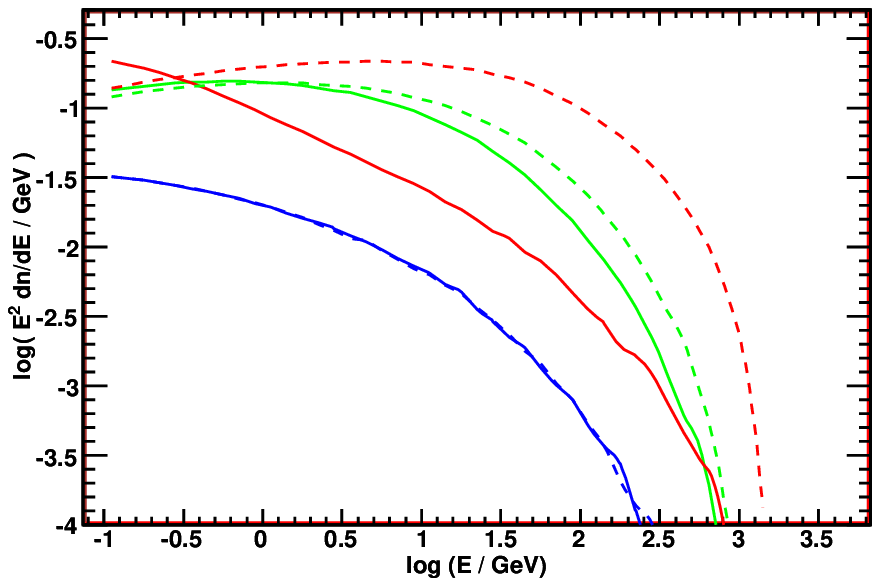}
\caption{As in Fig.~2, but for the equilibrium spectrum of electrons in the blob
derived from the observed spectrum of synchrotron photons during the flare from Mrk 501, observed during the 1997 April 16. The $\gamma$-ray spectra escaping from the blob without (with) the effects of cascading are shown by the dashed (full) curves. 
}
\label{fig:fullE}      
\end{figure*}

The $\gamma$-ray spectra escaping from the blob (in its rest frame) have been also calculated for the injection of electrons, with the equilibrium spectrum derived 
from the observations of the X-ray flare from Mrk 501 during the 1997 April 16. 
In Fig.~3 we show the variety of spectral shapes of escaping $\gamma$-rays, which are expected in such optically thick SSC model as a function of some parameters (full curves). 
For large Doppler factors the absorption of $\gamma$-rays is negligible and the cascade does not develop. Then,  $\gamma$-ray spectra peak  at large energies, provided that synchrotron losses of primary electrons do not dominate (e.g. blue curve in the left-upper figure). 
Note that in the observer frame the spectra are additionally shifted to higher energies due to the
relativistic beaming with the Doppler factor D. On the other hand, the $\gamma$-ray spectra have relatively low intensities if the blob is large (longer variability time scales and stronger magnetic fields inside the blob). 
In such a case, electrons loose energy mainly on synchrotron processes but density of synchrotron photons is on the low level (due to the large blob), preventing efficient production of IC $\gamma$-rays (see blue curve in the right-bottom figure).  
If the blob moves with large Doppler factors (with values closer to the upper bound derived from observations of superluminal motions in AGNs), then the dimension of the blob is larger and the optical depths for $\gamma$-rays lower. In such a case, the cascading effects can be neglected (for comparison see blue dashed and full curves). However, for low Doppler factors ($D<10$),
the cascading effects are important (compare the dashed and full red curves). As a result, the spectra of escaping $\gamma$-rays peak at $\sim$GeV energies (in the observer's frame), showing relatively low fluxes at TeV energies. In general, more compact blobs with stronger magnetic field produce
strong synchrotron X-ray flares without accompanying strong TeV $\gamma$-ray flares,
due to the internal absorption of $\gamma$-rays in the blob. 

\section{Conclusions}
\label{conclusions}

Our detailed calculations of the internal absorption of $\gamma$-rays confirm the conclusion reached in Bednarek \& Protheroe~(1999) that the absorption of $\gamma$-rays produced in SSC model in the April
1997 flare in Mrk 501 can be important. The best consistency with the observed $\gamma$-ray spectrum is obtained for the magnetic field of the order of $\sim$0.1 G and the Doppler factor in the range $10-15$ (see also Bednarek \& Protheroe~1999).

Based on the calculations of the radiation processes in terms of the optically thick SSC  model, we suggest the existence of a new class of the BL Lac
type AGNs. In the case of such BL Lacs, strong X-ray flares should not be accompanied by strong TeV $\gamma$-ray flares. This new type of $\gamma$-ray emitting BL Lacs should show simultaneous flares in hard X-rays and GeV $\gamma$-rays. We propose that future experiments, with improved sensitivity in the GeV energies, such as AGILE and GLAST, should search for such a type
of flares especially from the nearby BL Lacs from which TeV $\gamma$-ray emission
has not been observed so far. Such GeV flares in BL Lacs might  occur, provided that  very compact blobs move with relatively low Doppler factors.
In fact such a different types of flares from BL Lac type AGNs, i.e. hard X-ray flares accompanied by GeV $\gamma$-ray flares, might already sporadically occur in the TeV $\gamma$-ray BL Lacs. Due to  absorption of TeV $\gamma$-rays, the power can be reprocessed in the IC $e^\pm$ pair cascade process from the TeV to GeV energies. There are some signatures that such flares have been already observed, i.e. the so called orphan X-ray flare reported from Mrk 421 (Rebillot et al.~2006).

\begin{acknowledgements}
JS would like to thank the organizers of the Conference for partial financial support and (JS and WB) the anonymous referee for useful comments. This research is supported by the Polish Komitet Bada\'n Naukowych grant 1P03D01028.

\end{acknowledgements}



\end{document}